\documentclass[pre,twocolumn,superscriptaddress,floatfix,amssymb,showpacs]{revtex4-1}
\usepackage{enumerate}
\usepackage{times}
\usepackage{graphicx}
\usepackage{amsmath}
\usepackage{color}
\usepackage{natbib}

\usepackage[pdfpagemode=None,colorlinks=true,urlcolor=black,%
linkcolor=blue,citecolor=blue,pdfstartview=FitH]{hyperref}
\bibpunct{[}{]}{,}{n}{}{;}

\graphicspath{{fig/}} 
\begin{document}

\title{Sedimentation of Knotted Polymers}

\author{J. Piili}
\affiliation{Department of Biomedical Engineering and Computational Science, Aalto University, P.O. Box 12200, FI-00076 Aalto, Finland}
\author{D. Marenduzzo}
\affiliation{School of Physics, University of Edinburgh, Mayfield Road, Edinburgh EH9 3JZ, Scotland}
\author{K. Kaski}
\author{R. P. Linna}
\affiliation{Department of Biomedical Engineering and Computational Science, Aalto University, P.O. Box 12200, FI-00076 Aalto, Finland}

\begin{abstract}
We investigate the sedimentation of knotted polymers by means of stochastic rotation dynamics, a molecular dynamics algorithm that takes hydrodynamics fully into account. We show that the sedimentation coefficient $s$, related to the terminal velocity of the knotted polymers, increases linearly with the average crossing number $n_c$ of the corresponding ideal knot. To the best of our knowledge, this provides the first direct computational confirmation of this relation, postulated on the basis of experiments in Ref.~\cite{rybenkov_sedimentation_analysis}, for the case of sedimentation. Such a relation was previously shown to hold with simulations for knot electrophoresis. We also show that there is an accurate linear dependence of $s$ on the inverse of the radius of gyration $R_g^{-1}$, more specifically with the inverse of the $R_g$ component that is perpendicular to the direction along which the polymer sediments. When the polymer sediments in a slab, the walls affect the results appreciably. However, $R_g^{-1}$ remains to a good precision linearly dependent on $n_c$. Therefore, $R_g^{-1}$ is a good measure of a knot's complexity.
\end{abstract}

\pacs{87.15.A-,82.35.Lr,82.37.-j}

\maketitle
\section{Introduction}\label{sec:intro}
The topology of knots has been studied for almost two centuries. The earliest work dates back to the 19th century~\cite{silver,sossinsky_knots}, when P.~G. Tait first proposed the current classification of knotted loops based on the crossings observed in their two-dimensional projection. Since then, knot theory has been applied to a wide number of areas in physics and beyond: of particular relevance to our work are the applications to biophysics, as several biopolymers are observed to form knots. A particularly important case is that of DNA~\cite{stasiak_electrophoretic, vologodskii_sedimentation,bates_dna_topology,arsuaga_dna_knots_phage,marenduzzo_dna-dna_capsids,witz_dna_knots_supercoiling,enzo1}. DNA has been long known to routinely form simple knots in solutions~\cite{liu}, and it has been argued that specific enzymes must be at work within a cell to avoid excessive DNA knotting~\cite{magnasco}, which could be detrimental for the correct genome function. More recently, knots have been discovered also in the native states of proteins, where they are however quite rare~\cite{mansfield}. Knots are further present in synthetic polymers. Recently an organic trefoil knot was created synthetically by self assembly methods~\cite{ponnuswamy_discovery_organic_trefoil}.

DNA knots are even more common within bacteriophage heads~\cite{calendar,liu_knotted_dna_capsids}. Arsuaga {\it et al.} showed that the tight geometric confinement the genome is subjected to is a significant contributor to the knot formation~\cite{arsuaga_knotting_probability,arsuaga_dna_knots_phage}. Such bacteriophages are the ``viruses of bacteria'', and they work by releasing their DNA into the cytosol of their host. This release, or DNA ejection, is essentially entropically driven, as the DNA of phages is confined to essentially crystalline density, which is highly costly entropically. 
Also wildtype phages when burst open reveal a knotted double-stranded DNA. Modeling has shown that the spectrum of knots observed in the bacteriophage P4~\cite{arsuaga_dna_knots_phage} can be understood on the basis of simple polymer physics models that view DNA as a semiflexible and self-avoiding polymer, provided that an aligning interaction between contacting DNA segments is included~\cite{marenduzzo_dna-dna_capsids}.

The existence of knotted double-stranded DNA is by no means restricted to the inside of bacteriophages. Using a variety of enzymes, knotted DNA have even been generated in a test tube~\cite{brown}. They form {\it in vivo} in non-replicating cells~\cite{shishido1,shishido2,ishii} and during replication, see {\it e.g.}~\cite{olavarrieta1,olavarrieta2}. Very recently it was shown that topoisomerase IV is responsible for the knotting and unknotting of sister duplexes during DNA replication~\cite{lopez_topoisomerase_iv}.

As the knots are so frequent and important in the biophysics of DNA, there is clearly a demand for their easy identification. But how can one tell whether a DNA molecule, whose (hydrated) thickness is about 2.5 nm, is knotted, and which knot it forms? There is a large body of work on experimental gel electrophoresis paving the way for relating the DNA topology to migration velocity (see {\it e.g.}~\cite{levene_topoisomers} and the references therein). Motivated by their previous finding that knots with the same minimal crossing number comigrate on gels~\cite{dean}, Stasiak and co-workers presented in their seminal paper experimental results on DNA gel electrophoresis~\cite{stasiak_electrophoretic} and showed that the measured electrophoretic mobility $\nu$ of the DNA knots used in the experiments increases linearly with the average crossing numbers $n_c$ of the ideal forms (defined below) of these knots~\cite{stasiak_electrophoretic,vologodskii_sedimentation}. This dependence makes the identification of DNA knot topology much faster than e.g. by using electron microscopy methods; it also allows the determination of knotting probability in an ensemble of DNA molecules without the need to examine each of these singularly.

How can we compute $n_c$ in practice? To do so, we first need to deform a knot into its ideal form, which is defined as the one with the highest ratio of volume to surface area~\cite{katritch_geometry_knots,stasiak_ideal_knots,grosberg_flory-type_theory,moffatt_energy_spectrum}. In other words, this is a knot that is formed with as short a ropelength (polymer) as possible. The average crossing number, $n_c$, is defined as the average number of crossings of all possible two dimensional projections of the knot in its ideal form~\cite{katritch_geometry_knots}. Clearly, the larger $n_c$ becomes, the more complex is the corresponding knot. Knots are typically indexed according to their minimal number of crossings, $n_{\rm min}$, which is the minimum number of crossings over all projections -- therefore $n_c\ge n_{\rm min}$.\textit{i.e.} the number of crossings that cannot be opened without breaking the knot contour.

The reason why the electrophoretic mobility of knots increases with average crossing number, hence their complexity, can be understood with a simple physical argument. If a DNA knot is subject to a force $f$, for instance due to an electric field, then the (terminal) velocity it will reach can be estimated by the formula $f=\gamma v$, where $\gamma$ is an effective friction -- this is because the dynamics of DNA is highly overdamped in solution. For a sphere of radius $\rho$ the effective friction could be estimated with Stokes law as $6\pi\eta\rho$, where $\eta$ is the solvent viscosity. The typical ``size'' (radius) of a polymer is commonly measured by means of its radius of gyration $R_g$. If the knots have the same contour length, it is then intuitive that the more complicated ones will be more globular, hence smaller in shape: their terminal velocity, hence mobility, will then be larger. We note that for this argument hydrodynamic interactions are crucial -- without these, the friction of a polymer would scale, according to the Rouse model, with the number of beads rather than with the radius of gyration (which scales as $N^{0.588}$ for a self-avoiding polymer).


This line of reasoning is essentially the one in~\cite{stasiak_electrophoretic,rybenkov_sedimentation_analysis}, where the authors used a method to calculate the expected sedimentation coefficient $s$ of DNA molecules with a given topology~\cite{rybenkov_sedimentation_analysis}, and found that $s$ increases linearly with $n_c$. However, the original computation that employed the Kirkwood-Riseman approximation neglected the effect of flow on polymer conformation, and was essentially an estimate only valid at infinitesimally small forcing. To quantitatively establish the claim on the linear relationship between $\nu$ or $s$ and $n_c$ one needs to simulate the sedimentation of knotted polymers using a computational method that studies the molecular dynamics of the polymer in the presence of hydrodynamics and of a gravity field. This is what we set out to do in the present paper. As our work comprises the first numerical simulations of knot sedimentation including full hydrodynamic interactions, our results and framework may be used in the future to enhance accuracy in knot determination by sedimentation experiments. We also highlight the importance of boundaries, which significantly affect the sedimentation coefficients we record. Our results may be seen as complementary to the dynamic Monte-Carlo simulations in Ref.~\cite{delosrios_dietler_electrophoresis1,delosrios_dietler_electrophoresis2} which established the linearity of the electrophoretic motility with $n_c$ via direct simulations of the dynamics of a knot in a gel, under an electric field.

Our paper is organized as follows. In Sec.~\ref{sec:cm} we explain our computational model for polymer sedimentation: the model for the DNA, the method for simulating the dynamics and the used simulation geometries. In Sec.~\ref{sec:res} we present the results and conclude in Sec.~\ref{sec:con}.

\section{The Computational Model}\label{sec:cm}
\subsection{The polymer model}\label{sec:pm}
The polymer is modeled as a circular chain of pointlike beads with mass $m_b$. The adjacent pairs are connected by means of the finitely extensible nonlinear elastic (FENE) potential
\begin{align}\label{equ:fene}
U_F = - \frac{K}{2} r_{\rm max}^2 \ln{\left( 1 - \left(\frac{r}{r_{\rm max}} \right)^2 \right)}\;,\; r < r_{\rm max},
\end{align}
where $r$ is the length of the bond and $r_{\rm max} = 1.5\sigma$ is the maximum bond length. All bead pairs interact through a shifted truncated Lennard-Jones (LJ) potential,
\begin{align}\label{equ:lj}
U_{LJ} = \left\lbrace \begin{array}{ccl}
		4 \epsilon \left[\left(\frac{\sigma}{r_{ij}}\right)^{12} - \left(\frac{\sigma}{r_{ij}}\right)^6\ \right] + \epsilon &,& r_{ij} \leq  \sqrt[6]{2}\sigma\\
		0		&,& r_{ij} > \sqrt[6]{2}\sigma,
		\end{array}.
 \right.
\end{align}
where $r_{ij}$ is the distance between beads $i$ and $j$.
The potential parameters are chosen as $\sigma = 1.0,\; \epsilon = 1.0$, and $K = 30/\sigma^2$. This truncated repulsive Lennard-Jones potential models a good solvent.

\subsection{Polymer and solvent dynamics}\label{sec:srd}
In the modeled sedimentation process the polymer is immersed in a solvent and is driven by a constant (gravitational) force. This makes sedimentation an inherently non-equilibrium process. For such processes hydrodynamics typically plays a significant role. Hydrodynamic interactions  are taken into account by modeling the solvent using stochastic rotation dynamics (SRD), a computationally efficient Navier-Stokes integrator~\cite{malevanets_orig}. We use a hybrid version of the algorithm where the polymer follows Newton's dynamics and SRD is applied to both the solvent and the polymer. The polymer molecular dynamics is implemented using the standard velocity Verlet algorithm~\cite{frenkel_moldy}. The above described model has been successfully used in modeling elasticity and hydrodynamics of linear DNA molecules~\cite{riku08}.

The solvent consists of pointlike particles, each of mass $m_s$. One SRD step comprises two smaller steps: the streaming step and the collision step. In the streaming step the positions of solvent particles are updated as
\begin{align}
\vec{r}_i(t + \Delta t) = \vec{r}_i(t) + \vec{v}_i(t)\Delta t,
\end{align}
where $\vec{r}_i(t)$ and $\vec{v}_i(t)$ are the location and the velocity of the particle $i$ at time $t$, respectively, and $\Delta t$ is the SRD timestep. The whole simulation space is divided into a grid of cubic cells of equal size. In each cell the velocities of particles are updated in the collision step as 
\begin{align}\label{equ:srd_step}
\vec{v}_i(t+\Delta t) = \vec{v}_{\rm cm}(t) + \Omega \left[ \vec{v}_i(t) - \vec{v}_{\rm cm}(t) \right],
\end{align}
where $\vec{v}_{\rm cm}(t)$ is the center-of-mass velocity of the cell. Operator $\Omega$ is a rotation matrix whose rotation angle $\alpha$ is fixed but the rotation axis is chosen randomly in each timestep for each cell. The method conserves energy and momentum in each cell. In order for the method to maintain Galilean invariance the grid is shifted randomly at each step~\cite{ihle_srd}.

The polymer is coupled to the solvent in the collision step, Eq.~\eqref{equ:srd_step}. Here, the polymer beads are treated like the solvent particles. The resulting polymer beads' velocities are then used in the following velocity Verlet steps performing molecular dynamics (MD). MD  and SRD take turns so that after every 500 MD steps of timestep $\delta t = 0.002$ is performed a single SRD step of timestep $\Delta t = 1$. The other parameters were chosen as follows: the edges of the cubic cells are of unit length, the average density of solvent particles is 5 particles per unit cell, and the rotation angle $\alpha = 3\pi/4$. The mass of the solvent particles is $m_s = 4$ and the mass of the polymer beads is $m_b = 16$.

\subsection{The simulation geometry and polymer configurations}
The simulation space is a parallelepiped with periodic boundary conditions along the $y$ and $z$ directions for both the polymer and the solvent particles. The polymer is driven by a constant force in the negative $z$ direction. To model different experimental setups where sedimentation takes place in a large container and in a channel we run the simulations with either periodic boundary conditions also along the $x$ direction, or in the presence of two no-slip walls parallel to the $yz$ plane and normal to the $x$ direction. When the solvent particle hits the no-slip wall its momentum is reversed, so it is bounced back to the direction of incidence~\cite{lamura_multi-particle_cd_flow_circular}. In addition to this, for the polymer there is a repulsive Lennard-Jones potential similar to Eq.~\eqref{equ:lj} with a cutoff at distance $0.1$ (in simulation units) from the wall. The dimensions of the parallelepiped are $L_x \times L_y \times L_z = 50 \times 50 \times 300$ when the periodic boundary conditions are applied and $L_x \times L_y \times L_z = 5 \times 50 \times 300$ when the no-slip walls are present.

In each simulation we use a circular polymer with the number of beads $N_b = 215$. However, initial configurations differ since we examine polymers with different knot topologies. The different knot topologies are created with the \textit{Knotplot} program~\cite{knotplotPhD,knotplotUrl}. Individual realizations of the real conformations corresponding to a single knot topology were created by thermalizing the knots using different stochastic forces due thermal fluctuations. Due to this and the twist potential not included in the standard coarse-grained DNA model, our simulations address the case of relaxed - not supercoiled - knots. In electrophoresis the setup corresponds to the migration of knots driven by a low voltage. Snapshots of the initial configurations and of typical conformations (during sedimentation) are shown in Fig.~\ref{fig:cartoon}, for three different knot types.

On each polymer bead we apply a constant force $f = m_b a = 0.1$ in the negative $z$ direction, where $a$ is the acceleration. Once the polymer has reached its terminal velocity $v_t$ we calculate the time average of the sedimentation coefficient 
\begin{align}\label{equ:sedi}
 s = \frac{v_t}{a}
\end{align}
in the negative $z$ direction. Applying the constant force to the polymer induces momentum in the polymer which transfers to the solvent via the SRD collision step, Eq.~\eqref{equ:srd_step}. 

When periodic boundary conditions are applied in the $z$-direction the unphysical increase in the momentum of the fluid due to the body force has to be judiciously taken into account and compensated for. This we do by removing the momentum $I / N_s$ from every solvent particle at each streaming step. $I = f N_b \Delta t$ is the impulse applied to the polymer due to the constant force between SRD-steps. The same impulse is then divided equally to each of the $N_s$ solvent particles. The procedure preserves the local momenta in the fluid induced by the sedimenting polymer but removes the unphysical net momentum building up due to the periodic boundaries.

The no-slip walls instead naturally dissipate the induced energy. Hence, there is no need for momentum correction for the sedimentation simulations within the channel. The solvent is kept at a constant temperature by scaling the momenta so that the equipartition theorem always holds.

\section{Results}\label{sec:res}
\begin{figure}[t]
\includegraphics[width=0.70\linewidth]{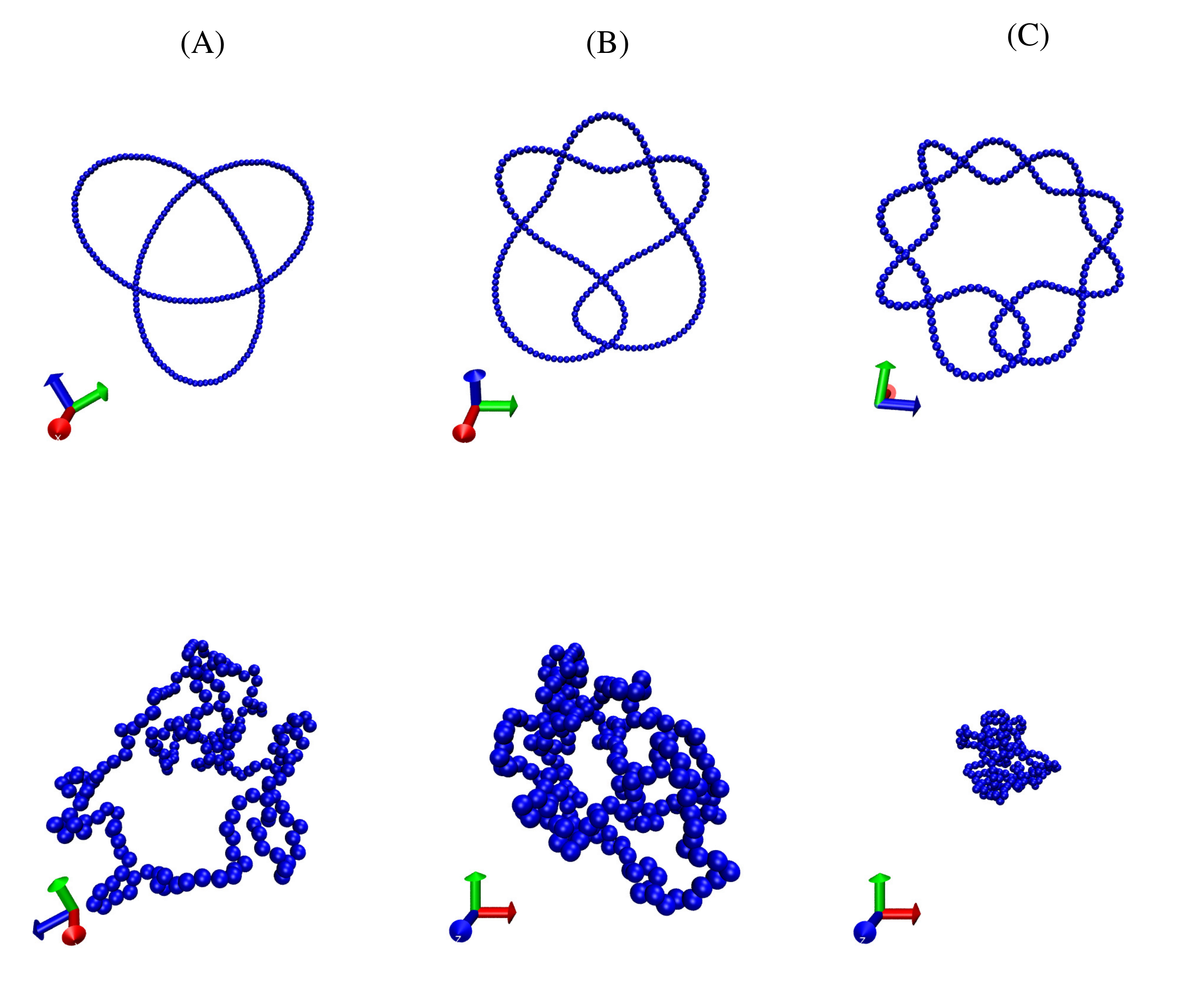}
\caption{(Color online) Initial configurations (top row) and typical configurations during sedimentation runs (bottom row) for three knots: trefoil ($3_1$, (A)), Stevedore's knot ($6_1$, (B)) and $10_1$ knot (C). These are three examples of twist knots, with the trefoil being a torus knot as well.}
\label{fig:cartoon}
\end{figure}
The sedimentation coefficient for each polymer knot topology was obtained by averaging over $55$ sedimentation simulations. In each simulation the sedimentation coefficient was averaged over $50000$ measurements. We computed the knot invariant \textit{Alexander polynomial} \cite{alexander_polynomial, rolfsen_knots_and_links} of the polymer after each run to verify that the knot was not lost due to phantom bond crossings during simulation. The knot topologies were sustained throughout the simulation runs, which is in accord with previous findings that the probability of phantom crossings is very small for polymers modeled by LJ and FENE potentials~\cite{kanaeda_diffusion_bond_crossing, kanaeda_universality_diffusion}.

The standard error of the mean was used as the estimate of error. In what follows, we denote an affine dependence of a variable $y$ on a variable $x$, \textit{i.e.} $y = Ax + B$, where $A$ and $B$ are constants, as $y \sim x$. Also, by 'linear dependence' we mean affine dependence. Stasiak and coworkers use linear dependence in this same manner. We give the formulas for the linear fits in figure captions along with the Pearson product-moment correlation coefficient $r_p$ (see {\it e.g.}~\cite{num_rec}) for comparing the linear relationships. These regression formulas will, of course, change when changing physical parameters such as the viscosity of the fluid.

\begin{figure}[t]
	\includegraphics[width=0.48\linewidth]{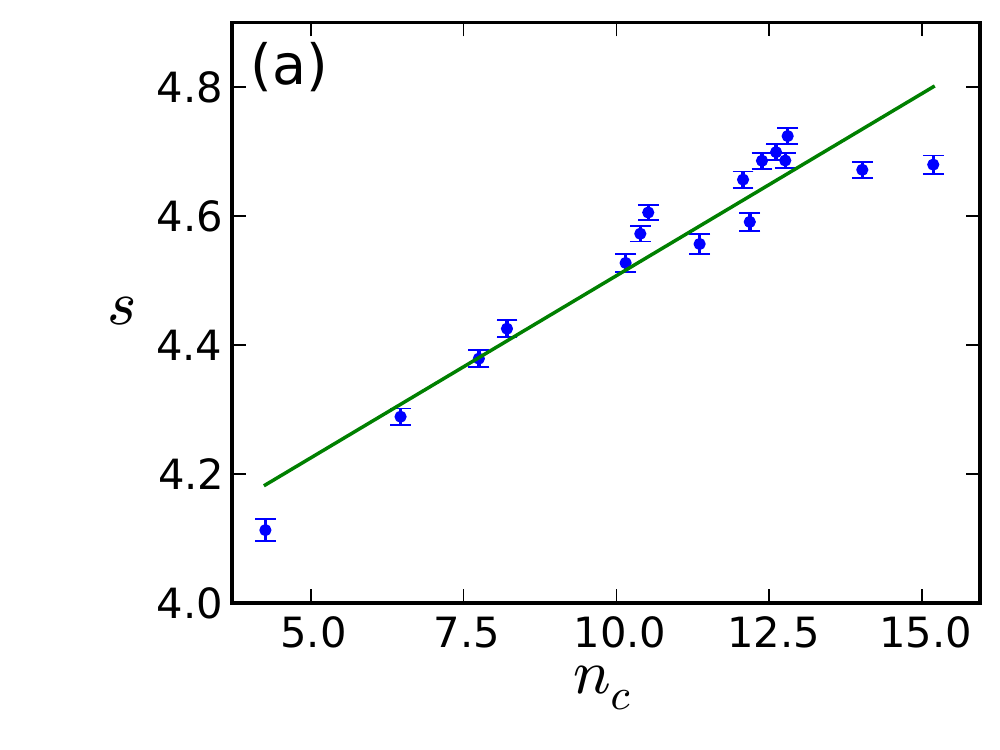}
	\includegraphics[width=0.48\linewidth]{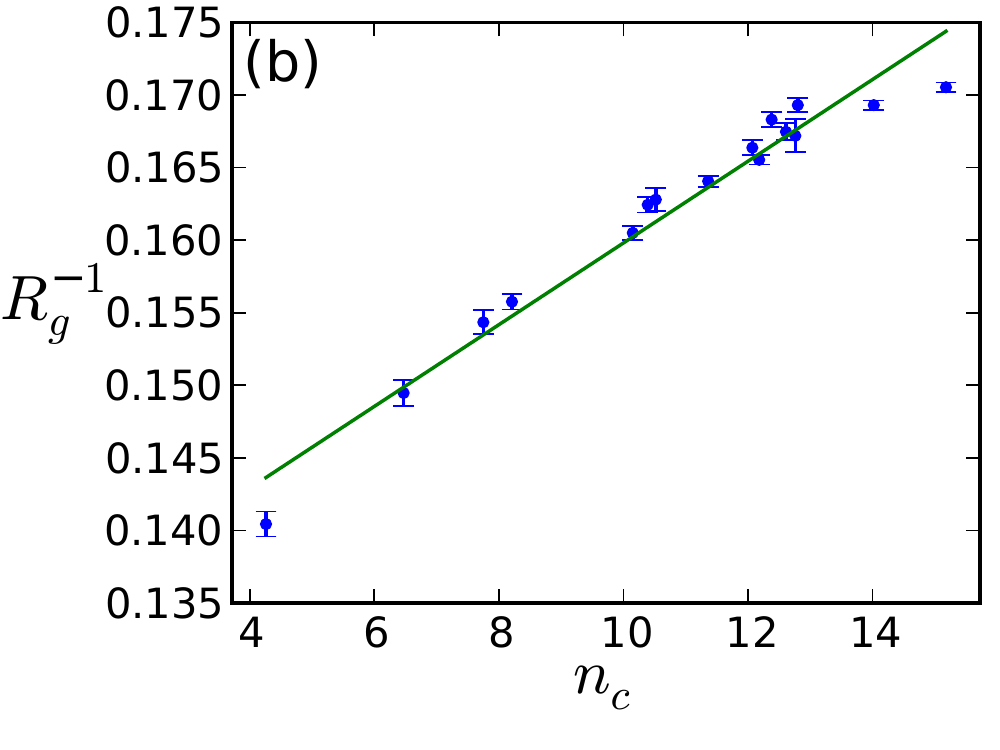}
	\caption{(Color online) Periodic boundary conditions. (a) Sedimentation coefficient $s$ and (b) the inverse of the radius of gyration $R_g^{-1}$ as a function of knot average crossing numbers $n_c$ for respective ideal knot conformations. Knots from left to right in all figures as function of $n_c$ are as follows: $3_1$, $4_1$, $5_1$, $5_2$, $6_1$, $6_2$, $6_3$, $7_1$, $7_3$, $7_2$, $7_5$, $7_4$, $7_6$, $7_7$, $8_1$, and $9_1$. Linear fits: (a) $s = 0.057 n_c + 3.9$, $r_p = 0.951$ (b) $R_g^{-1} = 0.0028 n_c + 0.13$, $r_p = 0.988$ }
	\label{fig:sedicrossing_gyratio_all_periodic}
\end{figure}

First we investigate the sedimentation of knots in the bulk (i.e. by using periodic boundary conditions rather than confining walls). Fig.~\ref{fig:cartoon} shows initial and typical configurations during sedimentation runs for three knots. In Fig.~\ref{fig:sedicrossing_gyratio_all_periodic}(a) the sedimentation coefficients $s$ for polymer knots are plotted against the average crossing numbers $n_c$ for their respective ideal forms. These crossing numbers are obtained from~\cite{stasiak_ideal_knots}. A very good linear relationship between $s$ and $n_c$ is obtained for the common DNA knot topologies that are also reported in the original papers of Stasiak and coworkers. For the more complicated, higher order knots, the dependence of $s$ on $n_c$ deviates somewhat from linearity. This aside, our direct molecular dynamic simulations including full hydrodynamic interactions confirm well that there should be a linear increase of $s$ with $n_c$, as proposed by Stasiak and coworkers, at least for relatively simple knots.

Stasiak and coworkers explained the linear relationship between $n_c$ and the knot migration speed as due to the fact that $n_c$ is directly proportional to the compactness of the knots. As a measure of molecular compactness they used the mean of inverse distances in a molecule~\cite{edwards}, $[R^{-1}] = \sum_{i,j (i \neq j)}^N \langle r_{ij}^{-1} \rangle$, where $\langle r_{ij}^{-1} \rangle$ is the mean reciprocal separation of segments $i$ and $j$~\cite{stasiak_electrophoretic}. The quantity $[R^{-1}]$ indeed appears in the approximate hydrodynamic theory used to compute sedimentation coefficients from polymer conformations~\cite{rybenkov_sedimentation_analysis} -- at the same time though this is hard to calculate in practice (see \textit{e.g.} ~\cite{miyake-mean-inverse}). Hence, it is not easy to show that $n_c$ increases linearly with $[R^{-1}]$ as would be required for the linear relationship between $n_c$ and the electrophoretic mobility $\nu$ to be proven rigorously.

The inverse radius of gyration $R_g^{-1}$ is also, by definition, a good measure of molecular compactness, and it can be readily and efficiently computed in simulations, as $R_g^2 = (1/N) \sum_{i=1}^N \langle (r_i-R_{cm})^2 \rangle$, where $R_{cm}$ is the center of mass. In Fig.~\ref{fig:sedicrossing_gyratio_all_periodic}(b) the inverse radii of gyration $R_g^{-1}$ measured for the sedimenting polymer knots are shown as a function of $n_c$. We obtain a good linear relation $R_g^{-1} \sim n_c$ for the knots investigated by Stasiak and coworkers. Interestingly, the slight deviations from the linear dependence $s \sim n_c$ show up in almost exactly the same way in $R_g^{-1} \sim n_c$. That is, also the relation $R_g^{-1} \sim n_c$ tends to break down for highly complex knots such as $8_1$ and $9_1$ -- just as the $s\sim n_c$ relation did.

\begin{figure}[t]
	\includegraphics[width=0.48\linewidth]{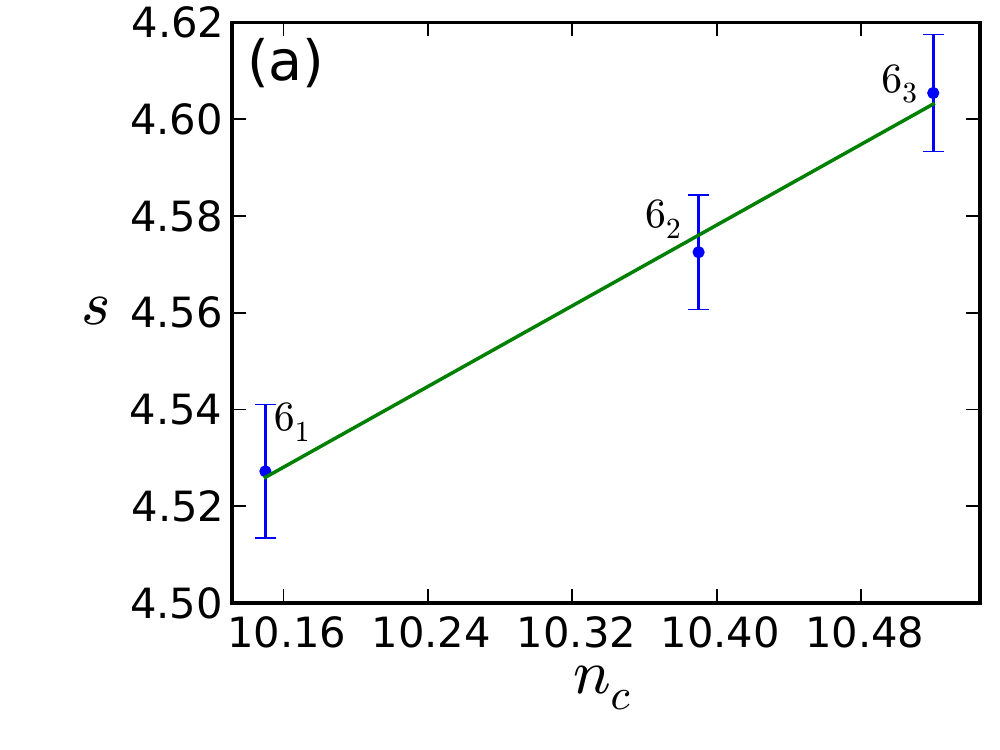}
	\includegraphics[width=0.48\linewidth]{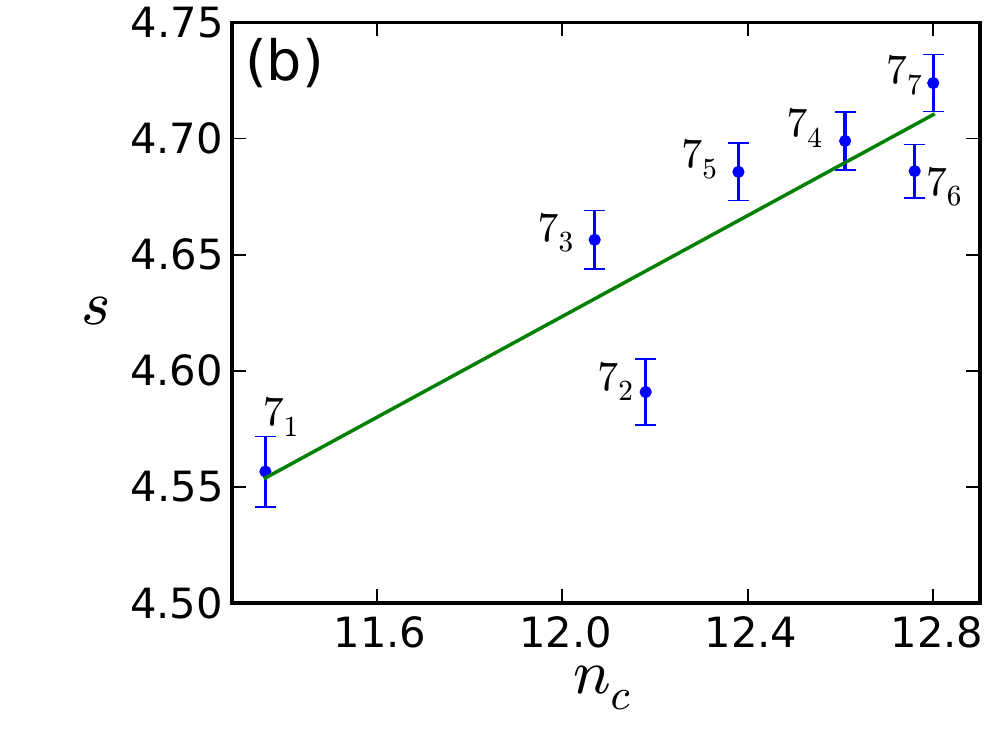}
	\caption{(Color online) Periodic boundary conditions. Sedimentation coefficient $s$ for knots with node numbers (a) $n_d = 6$ and (b) $n_d = 7$ as functions of knot average crossing number $n_c$. Within the knot family $n_d = 7$ the linear relation $s \sim n_c$ does not hold. Linear fits: (a) $s = 0.21 n_c + 2.4$, $r_p = 0.997$ (b) $ s = 0.11 n_c + 3.3$, $r_p = 0.894$.}
	\label{fig:sedicrossing_sub_periodic}
\end{figure}

By a careful look at the $s$ vs $n_c$ curves, one may note that the sedimentation coefficients for the knots within knot families with $6$ and $7$ minimal crossings deviate from the linear dependence $s \sim n_c$. The sedimentation coefficients $s$ obtained for all knots with $n_{\rm min} = 6$ and $n_{\rm min} = 7$ are shown in Fig.~\ref{fig:sedicrossing_sub_periodic}. For the knots $6_1$, $6_2$, and $6_3$, $s$ has a close to linear dependence on $n_c$ -- however, the slope differs from the one characterizing the general linear dependence on $n_c$ for all the knots. Within the knot family with $n_{\rm min} = 7$, $s$ violates linearity more significantly. For more complex knots (such as $8_1$ and $9_1$), deviations from linearity increase, as previously mentioned. All this is in good accord with the experimental observation that the linear relationship between the speed of electrophoretic migration speed and $n_c$ breaks down for complex knots~\cite{stark}.

\begin{figure}[t]
	\includegraphics[width=0.48\linewidth]{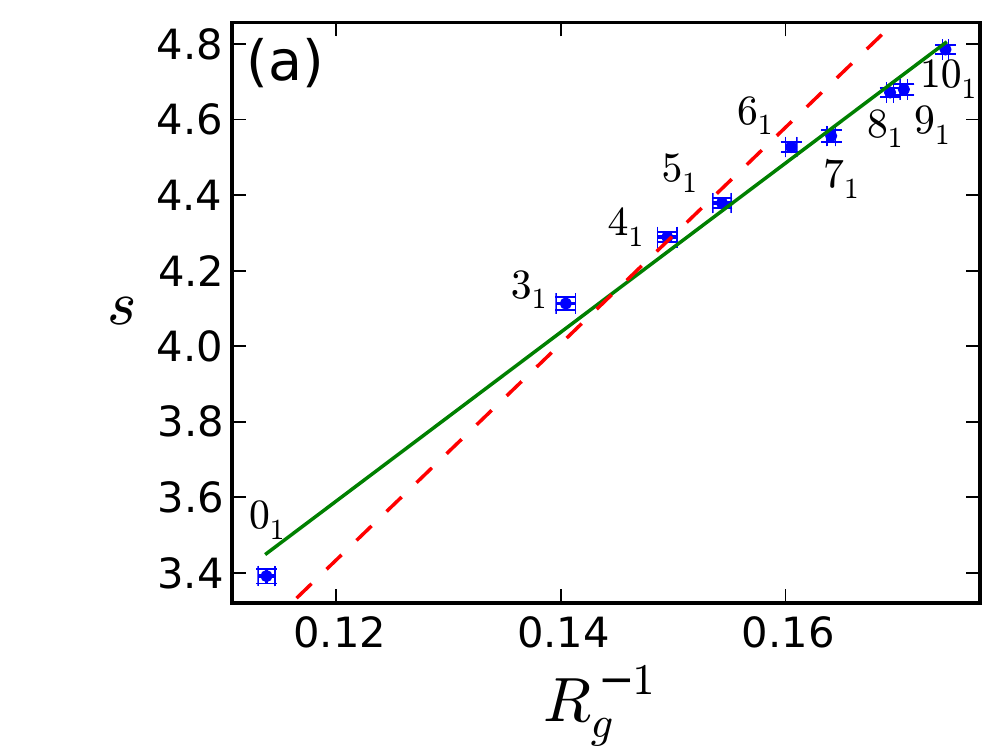}
	\includegraphics[width=0.48\linewidth]{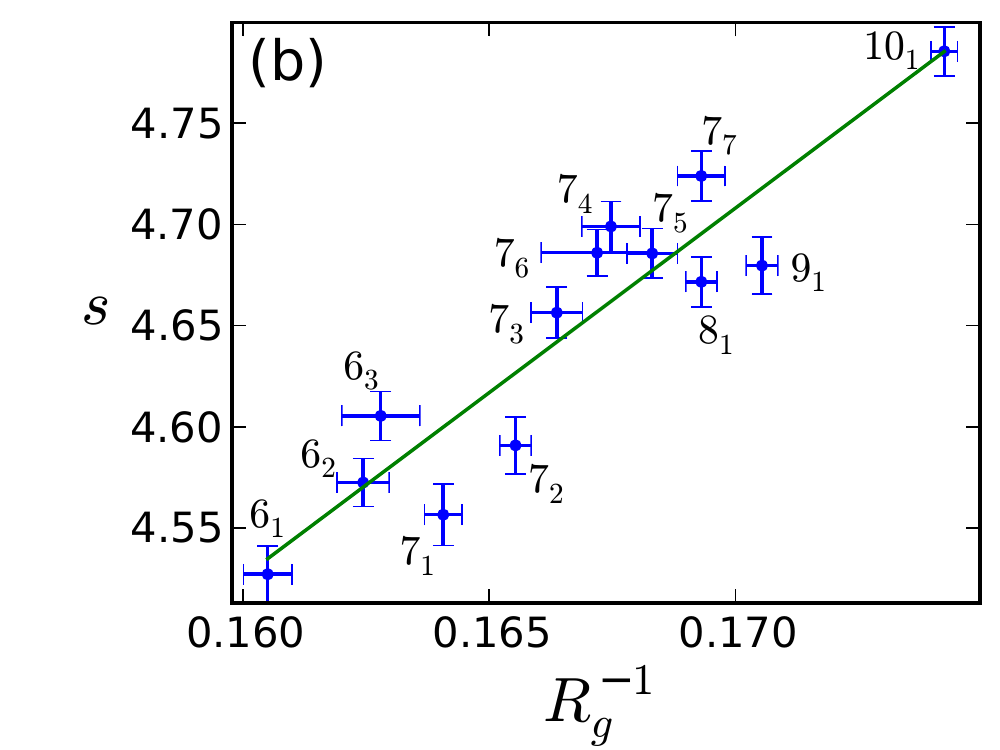}
	\includegraphics[width=0.48\linewidth]{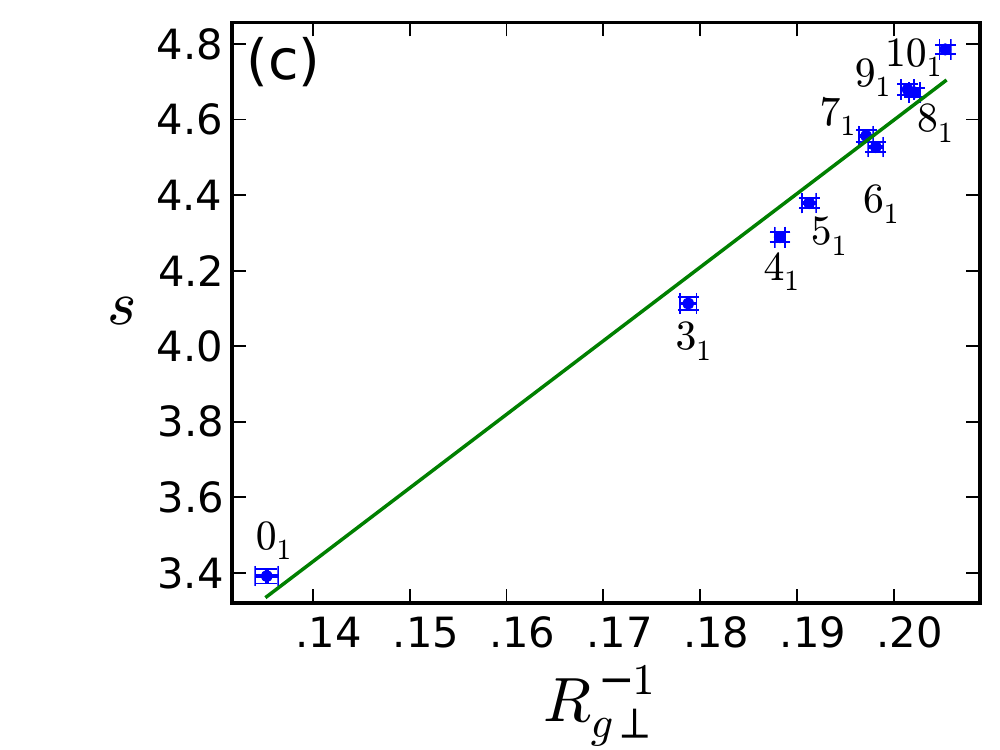}
	\includegraphics[width=0.48\linewidth]{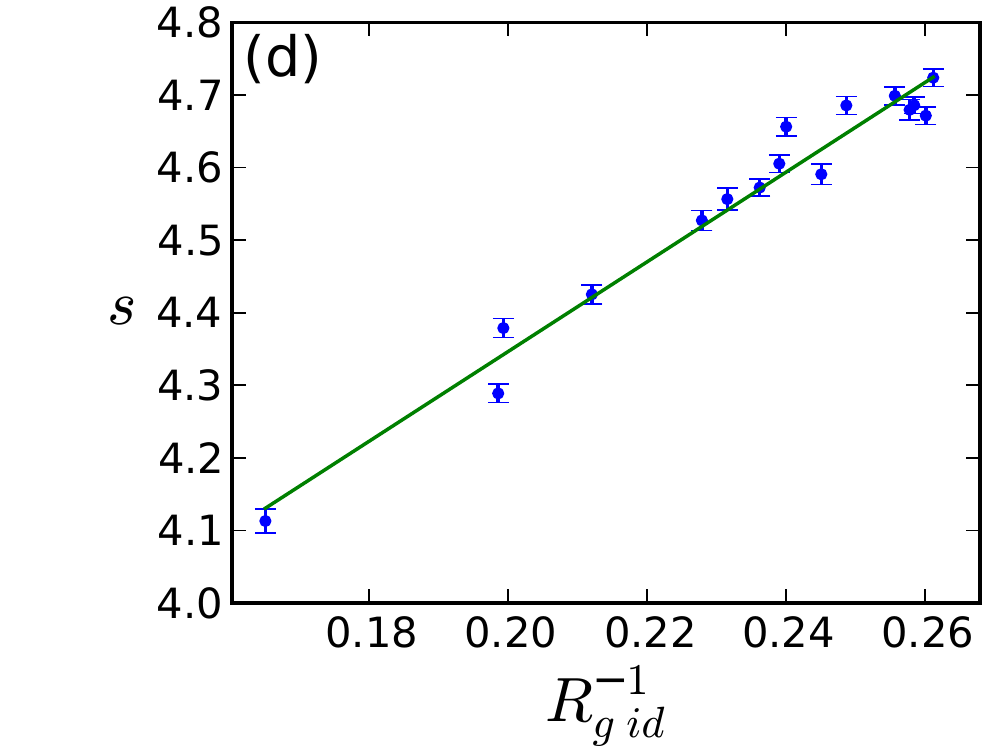}
	\caption{(Color online) Periodic boundary conditions. (a) Sedimentation coefficient $s$ as a function of the inverse radius of gyration $R_g^{-1}$. Only the knots with subscript 1 are depicted. Also the unknot $0_1$ and the knot $10_1$ are included. The dashed (red) line is the estimate for $s$ based on the Stokes' formula Eq.~(\ref{equ:stokes}), see text. (b) Magnification of Figure (a) where more complex knots of the families 6 and 7 are included. (c) $s$ as a function of the inverse of the perpendicular component of radius of gyration $R_{g \perp}^{-1}$. (d) Sedimentation coefficient as a function of the inverse of radius of gyration of the corresponding ideal knot configurations $R_{g\; id}^{-1}$. Linear fits: (a) $s = 22 R_g^{-1} + 0.91$, $r_p = 0.995$ (b) $s = 18 R_g^{-1} + 1.6$, $r_p = 0.926$ (c) $s = 19 R_{g\perp}^{-1} + 0.70$, $r_p = 0.990$ (d) $s = 6.2 R_{g\;id}^{-1} + 3.1$, $r_p = 0.983$. Knots in (d) from left to right: $3_1$, $4_1$, $5_1$, $5_2$, $6_1$, $7_1$, $6_2$, $6_3$, $7_3$, $7_2$, $7_5$, $7_4$, $9_1$, $7_6$, $8_1$, and $7_7$. (The ideal knot configurations were originally generated for publications~\cite{ashton_knot_tightening_gradient,rawdon_computers_ideal_knots}. Kindly provided by Eric Rawdon. Some ideal knot configurations can also be found at~\cite{kantarella_web}).}
	\label{fig:invgyratio_sedi_periodic}
\end{figure}

Fig.~\ref{fig:invgyratio_sedi_periodic}~(a) shows $s$ as a function of $R_g^{-1}$ for the first knot type (subscript $1$) of each knot family $n_{\rm min} \in \{0,3,4, \ldots ,10\}$. A very accurate linear relation with the Pearson product-moment correlation coefficient $r_p = 0.995$ is observed.
In other words, the linear relationship between $s$ and $R_g^{-1}$ (at least for sedimentation runs in the bulk) appears to hold more accurately than the one between $s$ and $n_c$. (That is because the dependencies of $s$ and $R_g^{-1}$ on $n_c$ show similar deviations from linearity, as observed above.) In Fig.~\ref{fig:invgyratio_sedi_periodic}~(b) $s$ vs $R_g^{-1}$ is shown for all the knots in the families with crossing numbers 6 and 7. For these more complex knots the linear relationship $s \sim R_g^{-1}$ is roughly that for all the knots, see Fig.~\ref{fig:invgyratio_sedi_periodic}~(a). In contrast, the deviation of $s \sim n_c$ for the more complex knots from $s \sim n_c$ for all the knots is clear, see Figs.~\ref{fig:sedicrossing_gyratio_all_periodic} (a),~\ref{fig:sedicrossing_sub_periodic} (a), and (b) and the related formulas for the linear fits.

The dependence $s \sim R_g^{-1}$ is in agreement with the simple physical argument given in the introduction; it has also been derived for DNA supercoils~\cite{odijk_sedimentation_of_dna_supercoils}, based on the Kirkwood-Riseman expression valid for linear Gaussian chains of radius of gyration $R_g$~\cite{kirkwood-riseman}. Using the Kirkwood-Riseman expression entails assuming that the sedimentation is non-draining, \textit{i.e.} that the motion of the solvent particles in the region of the polymer is largely suppressed (this is essentially the same approximation used in computing sedimentation coefficients in~\cite{rybenkov_sedimentation_analysis}). Consequently, backflow couplings between solvent flow and polymer conformations are completely neglected in that approach (unlike in our SRD simulations). 

It is instructive to compare the values of $s$ measured from our simulations with the result expected from the Stokes' formula for friction (see Introduction). For a spherical ball of radius $\rho$ the sedimentation coefficient would be
\begin{align}\label{equ:stokes}
 s = \frac{M}{6\pi\eta \rho}.
\end{align}
In Fig.~\ref{fig:invgyratio_sedi_periodic}~(a) we plot with the dashed line the sedimentation coefficient $s$ obtained using this formula when the radius $\rho$ is replaced by $\rho = 1.33 R_g$. Here, $M$ is the total mass of the polymer and $\eta = 4.78$ is the viscosity of the fluid estimated for SRD using formulas derived in Ref.~\cite{kikuchi_transport}. Hence, the correspondence between $s$ measured from the simulations and estimated from the Stokes' formula is reasonably good.

\begin{figure}[t]
	\includegraphics[width=0.48\linewidth]{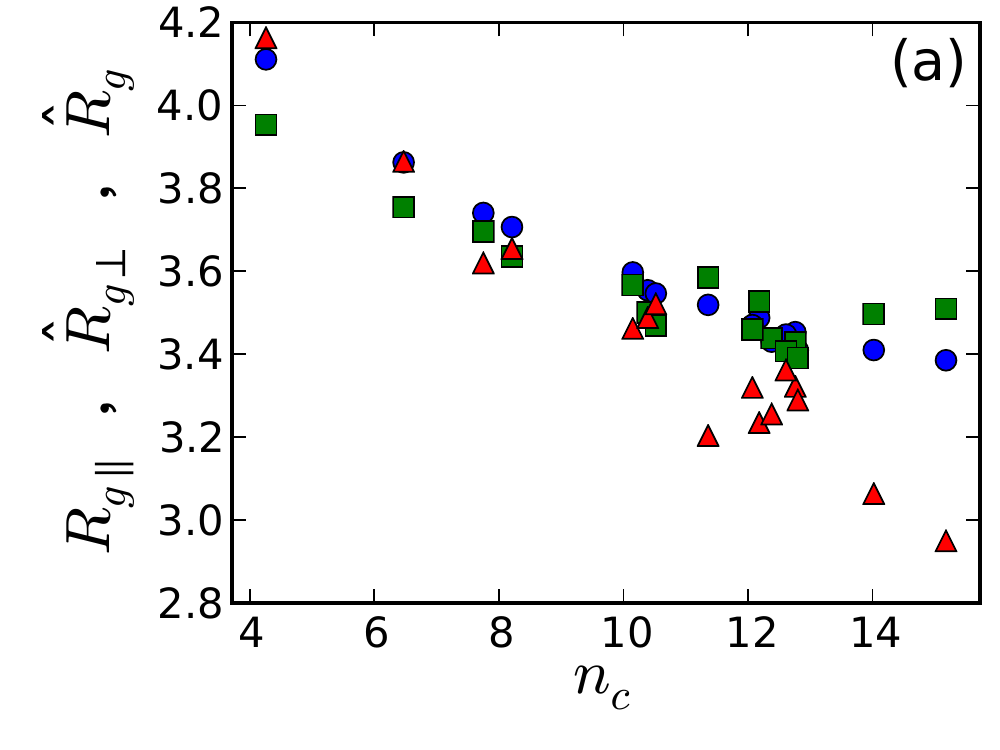}
	\includegraphics[width=0.48\linewidth]{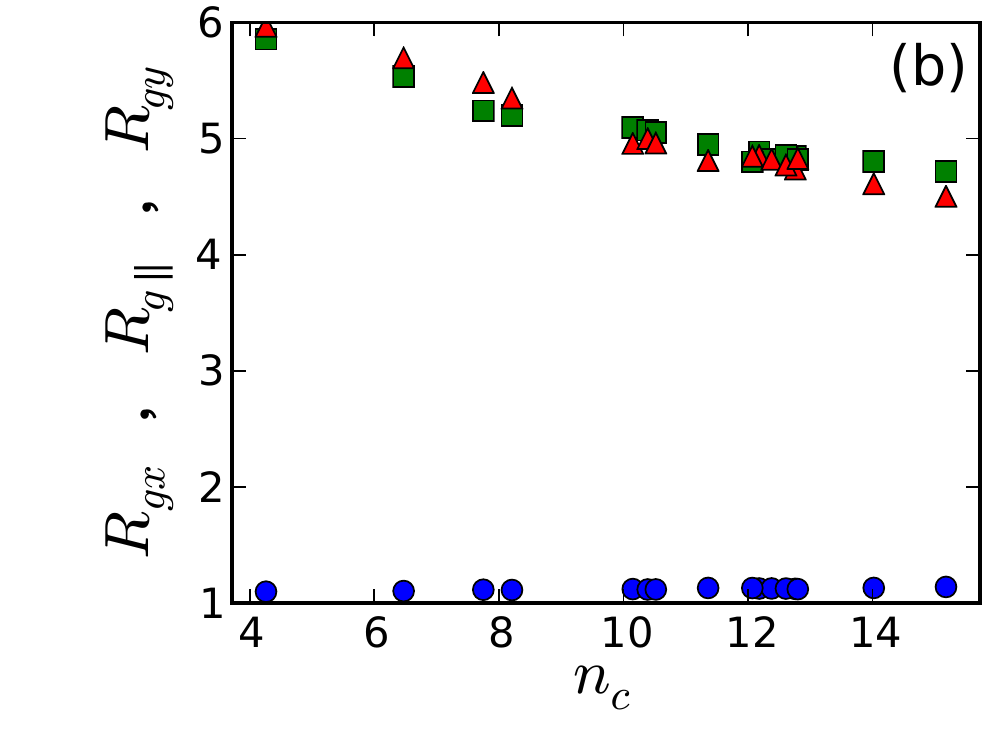}
	\caption{(Color online) (a) Different, appropriately normalized, components of radius of gyration with periodic boundary conditions. Circles: $\hat{R}_{g} = R_g/\sqrt{3}$, squares: $\hat{R}_{g \perp} = R_{g \perp}/\sqrt{2}$, triangles: $R_{g \parallel}$. \\(b) Different cartesian components of radius of gyration with slit walls. Circles: $R_{gx}$, squares: $R_{gy}$, triangles: $R_{g \parallel} = R_{gz}$.}
	\label{fig:gyratio_cross}
\end{figure}

Even though these simplified theories (Stokes drag and Kirkwood-Riseman non-draining approximations) work reasonably well, the effect of the velocity field on the polymer (backflow) cannot be wholly disregarded (nor can the near-field flow which is not captured by such approximations). 
The most relevant effect in our simulations which is not captured in the more simplified treatments is that the components of $R_g$ aligned, $R_{g \parallel}^2 = (1/N) \sum_i (r_{i,z} - R_{cm,z})^2$, and perpendicular, $R_{g \perp}^2 = (1/N) \sum_i [(r_{i,x}-R_{cm,x})^2 + (r_{i,y} - R_{cm,y})^2 ]$, to the direction of gravitational force differ appreciably, see Fig.~\ref{fig:gyratio_cross}~(a), at least for the values of accelaration $a$ used here. The dominant contribution to the linear dependence $s \sim R_g^{-1}$ comes from $s \sim R_{g \perp}^{-1}$, see Fig.~\ref{fig:invgyratio_sedi_periodic}~(c). Both $R_{g \parallel}^{-1}$ and $R_{g \perp}^{-1}$ depend linearly on $n_c$ to a good precision. However, the linear dependence $s \sim R_{g \parallel}^{-1}$ (not shown) is deteriorated. This in keeping with the friction being determined dominantly by the dimension $R_{g \perp}$. Interestingly, in comparison with the knots, the unknot ($0_1$) is very strongly elongated in the direction aligned with the gravitational force. Hence, in sedimentation the deformation of the knots differs markedly from linear and unknotted ring polymers.

To further asses the validity of the inverse of the radius of gyration as a measure the polymer knot complexity we plot the dependence of the measured sedimentation coefficients $s$ of the knots on the inverse radius of gyration of the topologically corresponding {\it ideal} knots of constant length $R_{g\ id}^{-1}$, see Fig.~\ref{fig:invgyratio_sedi_periodic}~(d). The obtained linear dependence of $s \sim R_{g\ id}^{-1}$ is more precise than $s \sim n_c$, see Fig.~\ref{fig:sedicrossing_gyratio_all_periodic}~(a).

\begin{figure}[t]
	\includegraphics[width=0.48\linewidth]{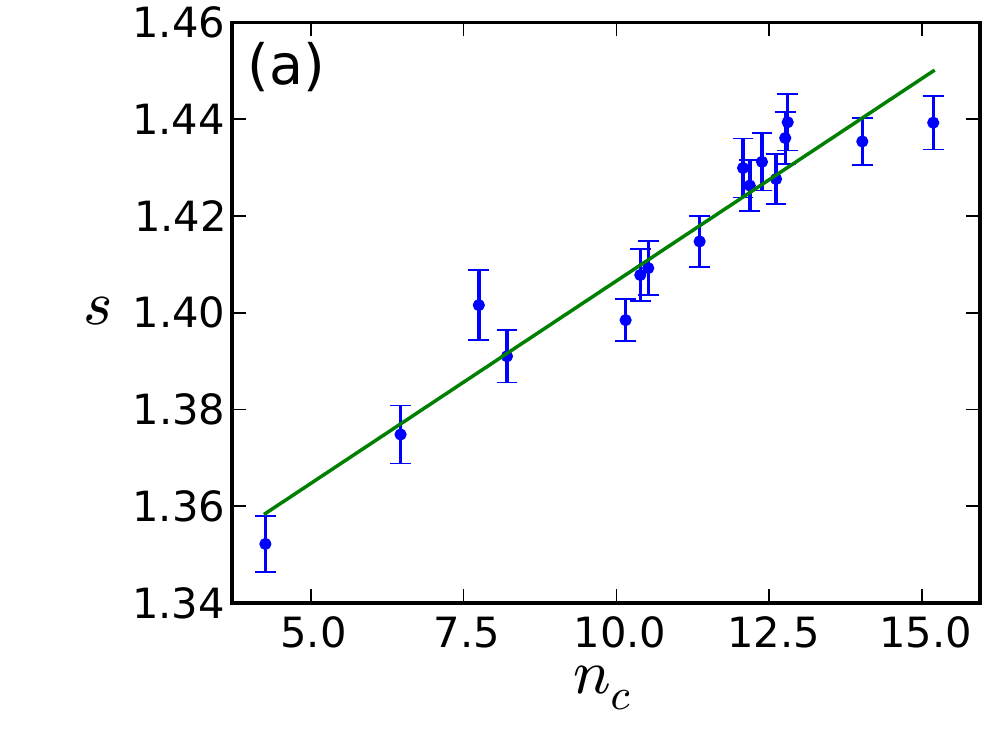}
	\includegraphics[width=0.48\linewidth]{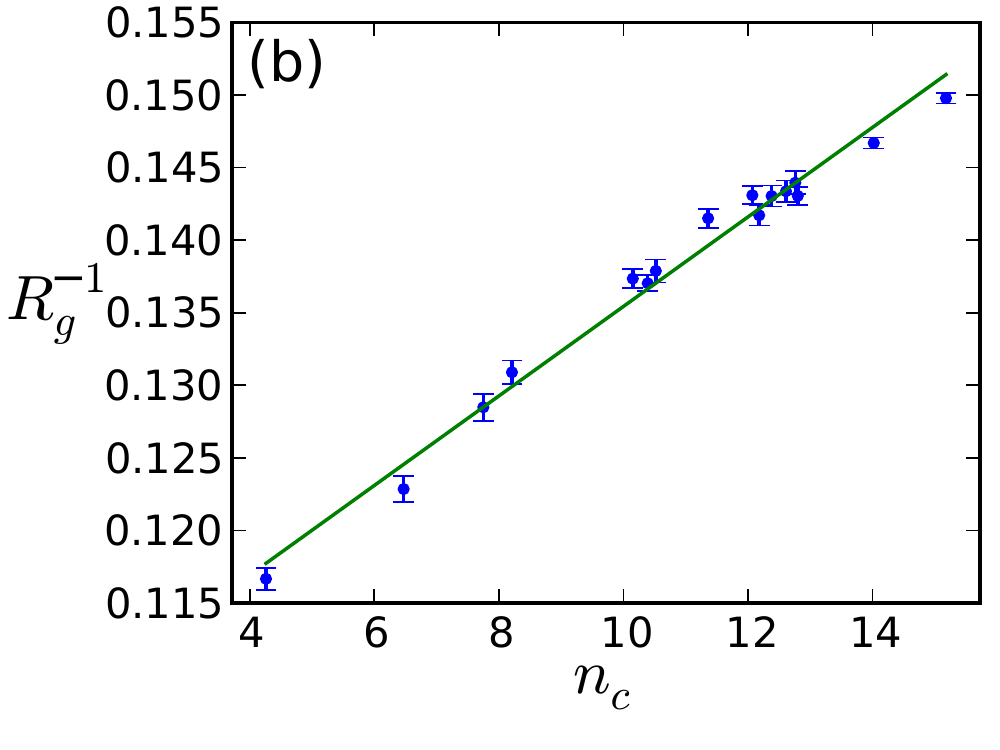}
	\caption{(Color online)  Slit walls. (a) Sedimentation coefficient $s$ as a function of average crossing number $n_c$. (b) The inverse of the average radius of gyration $R_g^{-1}$ as a function of knot average crossing number $n_c$. Fitted lines (a) $s = 0.0084 n_c + 1.3$, $r_p = 0.964$ (b) $R_g^{-1} = 0.0031 n_c + 0.10$, $r_p = 0.992$.}
	\label{fig:sedicrossing_gyratio_all_walls}
\end{figure}
\begin{figure}[t]
	\includegraphics[width=0.48\linewidth]{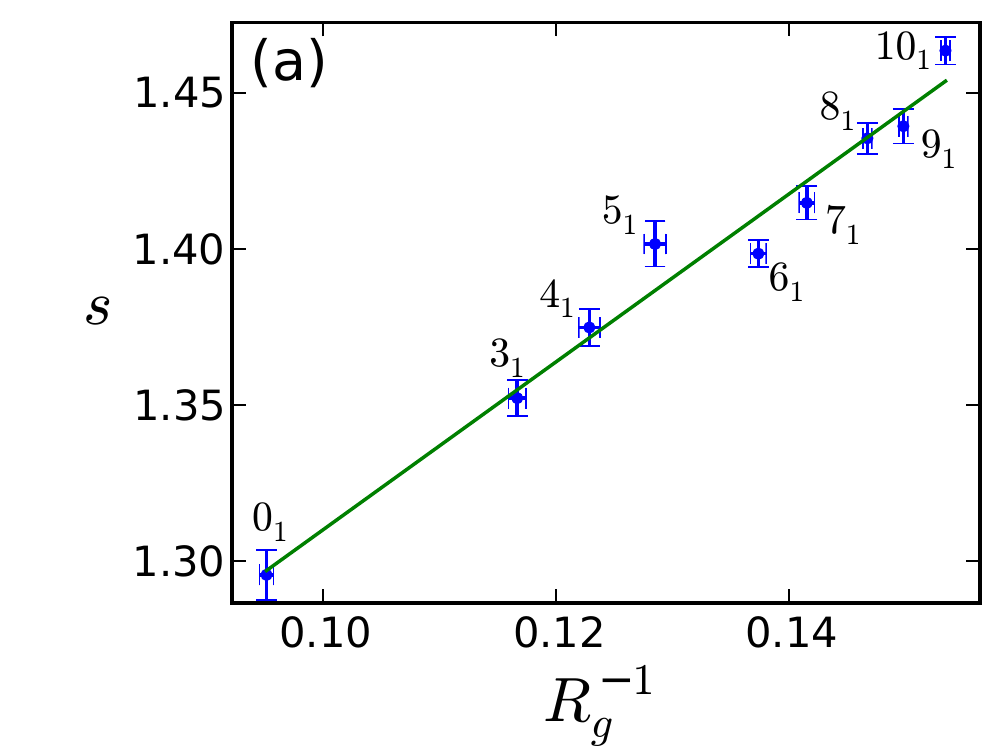}
	\includegraphics[width=0.48\linewidth]{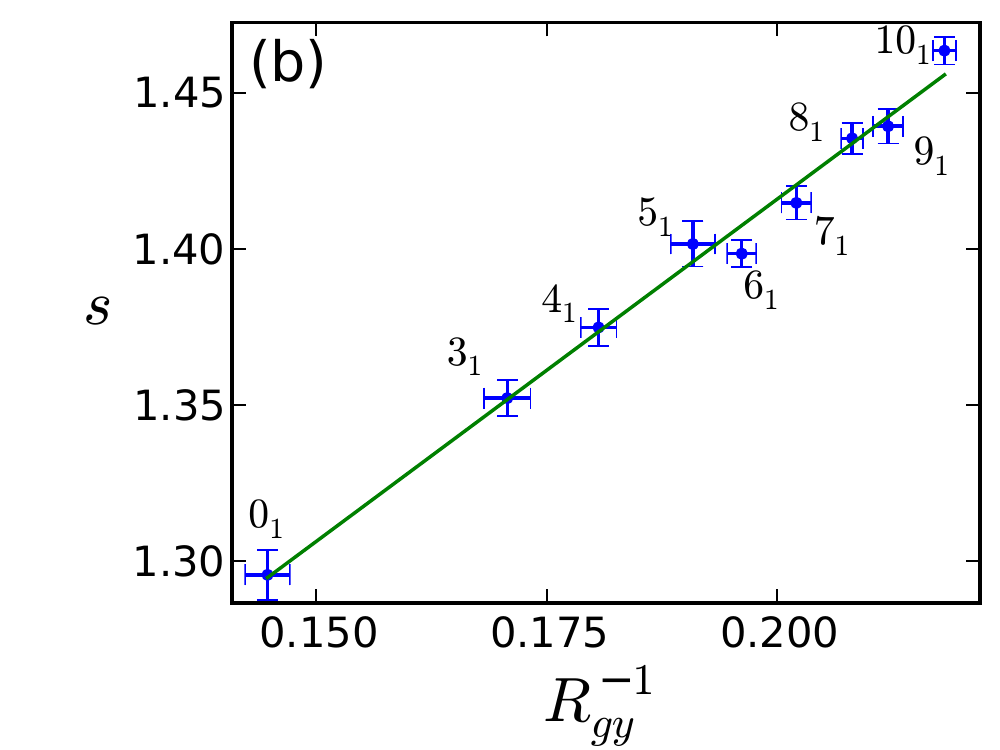}
	\caption{(Color online)  Results for sedimentation within a slab, with no-slip walls parallel to the direction of sedimentation. Sedimentation coefficient $s$ as a function of (a) the inverse of the radius of gyration $R_g$ and (b) the inverse the perpendicular component of the radius of gyration $R_{g \perp}$ (a) $s = 2.7 R_g^{-1} + 1.0$, $r_p = 0.987$ (b) $s = 2.2 R_{gy}^{-1} + 0.98$, $r_p = 0.995$.}
	\label{fig:invgyratio_sedi_walls}
\end{figure}
In order to consider the effect of boundaries, which is not necessarily negligible in the lab (or could be used to control the drag exerted by the fluid on the polymer), we also simulated sedimentation of knotted polymers between two no-slip walls. Due to the interaction of the fluid particles with the walls the friction felt by the polymer is enhanced, and as a result the polymer conformation is more elongated in the direction parallel to the walls. Hence, even for wide channels $R_{g \parallel}$ deviates from $R_{g \perp}$ more clearly than for the case of free solvent. When the channel is sufficiently narrow the polymer will be confined directly by the walls, see Fig.~\ref{fig:gyratio_cross}~(b). We show results for a channel of width $5$. For comparison, we measure $R_g \approx 7$ for the trefoil in free solvent. The linear relation between $s$ and $n_c$ is not as precise as in the bulk (periodic boundary conditions, no walls), see Fig.~\ref{fig:sedicrossing_gyratio_all_walls}~(a). At the same time the linear dependence of $R_g^{-1}$  on $n_c$ is still preserved, see Fig.~\ref{fig:sedicrossing_gyratio_all_walls}~(b). This suggests that $R_g^{-1}$ still reflects well the knot topology, but the sedimentation process is disturbed by the walls. 

Indeed, the precision of the relationship $s \sim R_g^{-1}$ for a polymer in the channel is deteriorated due to the walls, compare Figs.~\ref{fig:invgyratio_sedi_periodic}~(a) and \ref{fig:invgyratio_sedi_walls}~(a). Although the difference in Pearson product-moment correlation coefficients is not very large, the precision in identifying knot topologies through $s$ deteriorates under confinement (for example, consider the order of the knots $5_1$ and $6_1$). From Fig.~\ref{fig:gyratio_cross}~(b) it is clear that for polymers squeezed between the walls the component of the radius of gyration that is perpendicular to the walls $R_{gx}$ has no dependence on $n_c$. This happens already for wide channels (not shown). The linear dependence of $s$ on $R_{gy}^{-1}$, the component measured in the perpendicular direction on which periodic boundaries are applied, is preserved to a good precision, see Fig.~\ref{fig:invgyratio_sedi_walls}~(b), and contributes dominantly to the obtained $s \sim R_g^{-1}$, Fig.~\ref{fig:invgyratio_sedi_walls}~(a).

\section{Conclusion}\label{sec:con}
In conclusion, we have investigated sedimentation of knotted polymers using stochastic rotational dynamics, a computational model where hydrodynamics is taken fully into account, for the first time. Our motivation was to directly test the dependence of the sedimentation coefficient $s$ on the ideal average crossing number $n_c$, without making any assumptions about the segment distribution or hydrodynamic interactions. The linear dependence of $s$ on $n_c$ was seen to hold with fair to good accuracy for the knot topologies for which Stasiak and co-workers predicted this dependence. However, for some knots of higher complexity (such as 7-crossing number knots, $8_1$ and $9_1$), deviations from this dependence increase. This is in line with the experimental and theoretical results obtained for gel electrophoresis of knots~\cite{stark,delosrios_dietler_electrophoresis1,delosrios_dietler_electrophoresis2}. Our direct sedimentation simulations further justifies the argumentation commonly used in the literature to explain the observed linearity of $s$ on $n_c$ in gel electrophoresis experiments.

Our simulations show that the inverse radius of gyration of the polymer,  $R_g^{-1}$, is also proportional to the average crossing number $n_c$, and therefore provides another good measure of a knot's complexity. Interestingly, the deviations from linearity in the $R_g^{-1}$ vs $n_c$ curve pretty much mirror those in the $s$ vs $n_c$ curve. As a result, we observe that the linear relationship between $s$ and $R_g^{-1}$ is far more precise than the one between $s$ and $n_c$. As the solvent flow affects the knot conformations, and renders them more anisotropic, we monitored separately the dependence on the component of the radius of gyration parallel and perpendicular to the sedimentation direction: we found that it is the inverse of the latter which is more accurately
linearly proportional to $s$. 

Our fluctuating hydrodynamics simulations finally suggest that confinement, or the presence of boundaries, significantly affects the dependence of $s$ on $R_g^{-1}$, and may lead to a deterioration of the linearity between $s$ and $n_c$. The deviations from linearity in the $s$ vs $R_g^{-1}$ curves are due to $R_g$ component perpendicular to the walls losing all dependence on $n_c$.

We hope that our simulations will spur further results on knot sedimentation, aimed at verifying the deviations from linearity in the $s$ vs $n_c$ curve which can be observed due to confinement, or at high knot complexity. Eventually, finding a more precise relation for complex knots will lead to an improvement in the knot identification techniques when the average crossing number is high.

\begin{acknowledgments}
The computational resources of CSC-IT Centre for Science, Finland, and Aalto Science-IT project are acknowledged. We also thank Andrzej Stasiak, Cedric Weber, Giovanni Dietler and Eric Rawdon for useful e-mail communications.
\end{acknowledgments}

\bibliography{references.bib}

\end{document}